# From an attention economy to an ecology of attending. A manifesto.

**Authors:** Gunter Bombaerts* (1), Tom Hannes* (1), Martin Adam (2), Alessandra Aloisi (3), Joel Anderson (4), Lawrence Berger (5), Stefano Davide Bettera (6), Enrico Campo (7), Laura Candiotto (8), Silvia Caprioglio Panizza (8), Yves Citton (9), Diego D'Angelo (10), Matthew Dennis (1), Nathalie Depraz (11), Peter Doran (12), Wolfgang Drechsler (13), Bill Duane (14), William Edelglass (15), Iris Eisenberger (16), Beverley Foulks McGuire (17), Antony Fredriksson (8), Karamjit S. Gill (18), Peter D. Hershock (19), Soraj Hongladarom (20), Beth Jacobs (21), Gábor Karsai (22), Thomas Lennerfors (23), Jeanne Lim (24), Chien-Te Lin (25), Mark Losoncz (26), David Loy (27), Lavinia Marin (28), Bence Péter Marosán (29), Chiara Mascarello (30), David McMahan (31), Jin Y. Park (32), Nina Petek (33), Anna Puzio (34), Katrien Schaubroek (35), Jens Schlieter (36), Brian Schroeder (37), Shobhit Shakya (13), Juewei Shi (38), Elizaveta Solomonova (39), Francesco Tormen (30), Jitendra Uttam (40), Marieke Van Vugt (41), Sebastjan Vörös (42), Maren Wehrle (43), Galit Wellner (44), Jason M. Wirth (45), Olaf Witkowski (46), Apiradee Wongkitrungrueng (47), Dale S. Wright (48), Yutong Zheng (49)

*Affiliations:* (1) Eindhoven University of Technology, (2) University of Victoria, (3) University of Oxford, (4) Utrecht University, (5) Marist College, (6) European Buddhist Union, (7) University of Milan, (8) University of Pardubice, (9) University of Paris 8 (Vincennes Saint Denis), (10) Universität Würzburg, (11) University of Rouen, (12) Queens University Belfast, (13) Tallinn University of Technology, (14) Bill Duane and Associates, (15) Barre Center for Buddhist Studies, (16) University of Vienna, (17) University of North Carolina Wilmington, (18) University of Brighton, (19) East-West Center, (20) Chulalongkorn University, (21) independent writer, (22) Dharma Gate Buddhist College, (23) Uppsala University, (24) Being AI Corporation, (25) Tzu-Chi University, (26) University of Belgrade, (27) independent scholar, (28) TU Delft, (29) Budapest Business University, (30) Italian Buddhist Union Research Centre, (31) Franklin & Marshall College, (32) American University, (33) University of Ljubljana, (34) University of Twente, (35) University of Antwerp, (36) University of Bern, (37) Rochester Institute of Technology, (38) Nan Tien Institute, (39) McGill University, (40) Jawaharlal Nehru University, (41) University of Groningen, (42) University of Ljubljana, (43) Erasmus University Rotterdam, (44) Tel Aviv University, (45) Seattle University, (46) University of Tokyo, (47) Mahidol University, (48) Occidental College, (49) The Chinese University of Hong Kong)

*Co-first author, equal contribution and corresponding authors g.bombaerts@tue.nl and tom_hannes@yahoo.com

**Abstract**

As the signatories of this manifesto,

We denounce the attention economy as inhumane and a threat to our sociopolitical and ecological well-being.

We endorse policymakers' efforts to address the negative consequences of the attention economy's technology, but add that these approaches are often limited in their criticism of the systemic context of human attention.

Starting from Buddhist philosophy, we advocate a broader approach: an 'ecology of attending,' that centers on conceptualizing, designing, and using attention (1) in an embedded way and (2) focused on the alleviating of suffering.

With 'embedded' we mean that attention is not a neutral, isolated mechanism but a meaning-engendering part of an 'ecology' of bodily, sociotechnical and moral frameworks.

With 'focused on the alleviation of suffering' we explicitly move away from the (often implicit) conception of attention as a tool for gratifying desires.

We analyze existing inquiries in these directions and urge that they be intensified and integrated.

As to the design and function of our technological environment, we propose three questions for further research: How can technology help to acknowledge us as 'ecological' beings, rather than as self-sufficient individuals? How can technology help to raise awareness of our moral framework? And how can technology increase the conditions for 'attending' to the alleviation of suffering, by substituting our covert self-driven moral framework with an ecologically attending one?

We believe in the urgency of transforming the inhumane sociotechnical systems of the attention economy into a humane ecology of attending, and in our ability to contribute to it.

# 1. Introduction

The attention economy is the business-inspired approach that uses human attention as a commodity, exploiting impulsive reactivity via attention-grabbing designs/mechanisms. Its technology, with the exponential development of AI-applications, such as short-form social media content, brain-machine interfaces, eye-tracking, multimodal AI, and many more, add to the immersive impact of the attention economy. Attention is the new frontier of the capitalist process of enclosure and commodification, enabling economic and commercial logics to reach deep into our very subjectivities. There is an ongoing convergence of corporate and political acceleration in the use of techniques, including AI, to colonize and instrumentalize individual and collective experiences of attention.

In parallel with the attention economy's expansion, societal critique on its technology have increased exponentially too. Research has clearly shown the technology of the attention economy to be disruptive to the public-private balance, decreasing mental well-being worldwide, handing over power to a few high-tech companies, amplifying polarization and extreme political ideologies and contributing to global political instability (Gardenier et al., 2024; Mark, 2023).

These critiques, based on the defense of individual human rights, are already well-established and indispensable. In agreement with them, we join the voices warning of the inhumane effects of the current attention economy and its AI-applications. In section two, we analyze the way attention is conceived in the business-oriented approach of attention economy and in the lawmaker's attempts to curb its harmful effects. Even though these measures are undeniably important, we argue that they are merely defensive in character and focus too much on minimizing the negative aspects for individual users, partially by protecting privacy, limiting the impact on health, or blocking extreme online political messages.  In the third section of this article, we advance the debate around counteracting the attention economy by drawing inspiration from Buddhist philosophy. This angle has been less developed in this journal. Attention is a major key element in Buddhist philosophy and practice, and its insights already have been successfully applied to the study of societal aspects of AI (Hershock, 2021; Hongladarom, 2011; McGuire, 2016; Lin, 2023; Zheng, 2024; Tormen, 2023). Our criticism of the attention economy can also benefit from comparing a popular view of Buddhist attention as 'non-judgmental letting go' with a more integrated Buddhist view of attention as part of a moral 'ecology' (in the sense of embedded relationality) and aimed at 'attending' (as committed action to the alleviation of avoidable suffering) rather than at personal gratification (Hannes, 2023; Hadar & Ergas, 2019; Hermann, 2023).

In particular we draw inspiration from two aspects of Buddhist views on attention: the embedding of attention in the whole of an individual's existence, and the function of attention to alleviate suffering in the world. We will do so in order to develop a scaffolding for a more fundamental resistance against the attention economy in the fourth section. Several scholars have described attention as embedded in body, community, morality, epistemology, technology, or time. Others have developed a view on attention as a process or a constitutive practice, with a strong care for oneself (asking attention) or giving attention (to attend to). The Buddhist view on attention as serving to alleviate suffering is also put forward in technical, moral, sociological, communicational, legal, economic, business, and political approaches. However, we find that there is no coherence between these approaches. We make a plea to integrate these systemic approaches in order to describe the fundamental role of attention in AI-applications and developments and to enable a discussion that would move us from an attention economy towards an ecology of attending.

## 2. The attention economy and its liberal critics

This section describes two views on attention, each based on one core document taken as a token for a particular kind of discourse. We acknowledge that a single document cannot capture the entire richness of a particular discourse. For attention seen as an economic commodity (2.1), we refer to the seminal work "The Attention Economy: Understanding the New Currency of Business" by Davenport and Beck (2001). For attention seen as a personal right (2.2), we use the "European Parliament resolution of 12 December 2023 on the addictive design of online services and consumer protection in the EU single market (2023/2043(INI))". (European Parliament, 2023)

### *2.1. Attention as an economic commodity*

The term 'attention economy' refers to the business-marketing-economic system that treats human attention – which is a complex phenomenon, an embodied practice of living in a shared world – as a commodity, an economic resource with substantial fungibility or exchangeability. The attention economy does not only simply *use* attention as a commodity or a resource, it *reduces* it as such. Davenport and Beck define attention as "focused mental engagement on a particular item of information. Items come into our awareness, we attend to a particular item, and then we decide whether to act." (2001, p. 20)

The technological applications of the attention economy are based on simple but ultra-efficient behavior-change marketing models. They create ease of acting while minimizing effort, elicit psychological motivation to pay attention, and promise fulfilling rewards that, however, leave the users wanting more (Ibid., p. 24). In order to harvest users' attention and agency, they are reduced as much as possible to quantifiable impulsive reactions to AI-driven systems of enticement. (Ibid., p. 53) Emotions and motivations are considered mere means for grabbing attention. When considering mental states, there is a singular focus on subcortical reward learning systems, bypassing control by frontal regions. Davenport and Beck conclude: "Does this mean that to manage attention, business people should start centering all interactions around their associate's interest in sex, violence, and food? Not necessarily, though we are not ruling it out." (Ibid., p. 64)

In this system, in which attention is currency, data is the new gold. AI-systems are deployed to maximize the commodity's harvest and increase the companies' impact. Critical thinking about the attention economy system must be avoided, and proof of its damaging effects must be dodged. (Ibid., p. 9) The influence and impact of the attention economy system are presented as morally neutral, or as Davenport and Beck write: "Derber is undoubtedly right that people are narcissistic, and probably right that it [*authors: influencing attention*] is destructive. However, we are both realists, not social critics. How can an enterprising individual take advantage of the narcissistic nature of his or her peers?". (Ibid., p. 68)

### 2.2.-Attention as a personal right – a liberal critique

Despite its self-acclaimed neutral or 'realistic' stance, the huge success of the attention economy comes with a devastating effect on public and private welfare. Mainstream attention research illustrates its negative consequences, such as disrupting the public-private balance, decreasing mental well-being worldwide, the shifting and clustering power to a few high-tech companies, global political instability, and so on. (Gardenier et al., 2024; Mark, 2023) This has led to a widespread criticism of the side-effects of the attention economic system. Up to now, most critiques of the attention economy have aimed to minimize the diminishing of individual autonomy. Consumers "should not be undermined by traders' commercial practices" (European Parliament, 2023, §8) but should be offered "real choice and autonomy". (Ibid., §10) The critiques aim to limit impacts on physical and mental health, as well as privacy. (Ibid., §B, G, 3)

This essentially liberal perspective, conceived as the belief that it is the aim of politics to preserve individual rights and to maximize freedom of choice, considers attention not merely as an economic commodity but as a fundamental individual human right. Consumer protection legislation claims it "will reverse the negative trends that have been weakening consumers' position and reducing consumers' rights in a world dominated by digital technologies" (Ibid., §1, O) and that "policy initiatives and industry standards on safety by design in digital services and products can foster compliance with children's rights." (Ibid., §13) Some aspects of digital services are explicitly seen as positive: increased efficiency, connectedness, accessibility, and leisure, the possibility to connect, learn about, and appreciate different perspectives and worldviews, build knowledge and explore areas of interest, become more productive, exercise more, or solve specific problems. (Ibid., §B)

Although turning attention seeking features off-by-design is seen as crucial (Ibid., § O,10,13), educational guidelines, prevention plans, and awareness-raising campaigns "should promote self-control strategies to help individuals develop safer online behaviors and new healthy habits." (Ibid., §11) The individual mind is regarded as important; emotions are seen as signals of individual well-being. (Ibid., §D) Self-control plays an essential role. (Ibid., §D,G,I,J,K,O,8,9,11) So, whereas in the attention economy, the core aspect of attention is considered emotional reactivity, here, the focus is on autonomous individual agency.

The relevance and benefits of digital services and AI-systems are not questioned as such. Here too, data and information transfers are presented as essentially neutral. (Ibid., §A) Yet, as the liberal critique points out, digital services can create "power imbalances and digital asymmetry." (Ibid., §K) Therefore digital fairness is understood in terms of the proportionality of business and consumer rights (Ibid., §1, 5, 6), and users' efforts should be proportional and in line with their individual autonomous choices.

In this view, attention remains a resource to be exchanged, this time not as company property but as a right of the individual user. Companies are blamed for "capturing", "retaining" (Ibid., §A), or "regaining" (Ibid., §L) users' attention, so users should be protected against abuse.

But the elements of the AI-based systems (data, design, AI-algorithm, interactions, dark patterns in interface design…) are not conceptualized in such a way that is critical of the system itself. In line with green-washing, we could call this attention-washing. Such attention economy critiques are well-established and indispensable, yet at the same time, they remain limited to a primarily liberal approach focusing on minimizing the negative aspects for autonomous individuals in the negotiation of rights.

## 3. Buddhist inspiration for a wider criticism

In order to facilitate a broader approach to attention and to develop a positive alternative to the attention economy's technology, we draw inspiration from Buddhist philosophy. This requires us to see past the cliché of Buddhist attention practices as they are often presented in the West. Popular parlance tends to reduce Buddhism to its meditation practices, its meditation practices to 'mindfulness', and 'mindfulness' to nonjudgmental attention that lets go of all reactivity. (Bishop, 2004) This mindfulness is said to allow practitioners to see things as they are, free from outer influences or inner compulsions. (McMahan, 2008, 2023, p. 6)

Understood in terms of Davenport and Beck's model, the merit of this kind of mindfulness is that its practitioners linger longer in the first self-awareness-phase (e.g. Mascarello, 2018) so that they are more free to attend to what they want, and decide to act autonomously. In that way they are less reactive to the prompts of intrusive marketing strategies and AI-systems. This difference indeed matters.

But there is far more at play in Buddhist views of attention. For the reduction of Buddhism to mindfulness as nonjudgmental awareness obscures a great deal of Buddhism's pragmatics and, indeed, a great deal of our reality as human beings. The nature and role of attention in Buddhism is a far more complex and subtle matter than this article will be able to convey (Ganeri, 2020; Vörös, 2016). In the most technical considerations of early Buddhism, attention is a universal mental factor, meaning that it is a quality accompanying every moment of active consciousness. It is understood literally as "making in the mind" (*manisakara*[1]) (Bodhi 2000, p. 81) and is considered the force of the mind adverting towards an even slightly impinging object (Jacobs, 2017). This being said, there are a great many kinds of Buddhism. But our theme here is not 'Buddhist attention' as such. Our focus here is on two specific aspects of the way attention is presented in Buddhist sources, in order to broaden the criticism of the attention economy.

The first aspect is that attention is a practice that cannot be considered as neutral or in isolation, but needs to be seen as embedded. It is part of an 'ecology' of moral and epistemic frameworks. The second aspect focuses the Buddhist 'resolve' (*saṃkalpa*) to use attention for alleviating suffering in the world rather than as a tool for personal gratification. Next, we explain in more detail these two aspects.

---

[1] We mention the relevant concepts in Pāli for interested readers.

### 3.1. Dhammacakka: an ecology of attention

Davenport's and Beck's model represent attention as an isolated non-moral faculty in the AI-supported attention economy. Buddhism embeds attention within a wider context of practices. The earliest rendition of the Buddhist way is the so-called 'eightfold path'. Mindfulness indeed features in it as the seventh of eight domains of practice. All eight of them need to function in harmony in order to be functional, just as the iconic 'wheel of the teaching' (fig 1) needs its eight spokes to be attuned to each other for the wheel to turn properly. Early in its history Buddhist traditions grouped the eight aspects into three 'trainings' (Gombrich, 2009, p. 109). The training of wisdom consists of 'right view' and 'right resolve'. Moral training contains 'right speech', 'right action' and 'right livelihood'. And 'right effort', right mindfulness' and 'right concentration' form the training of the mind or meditation.

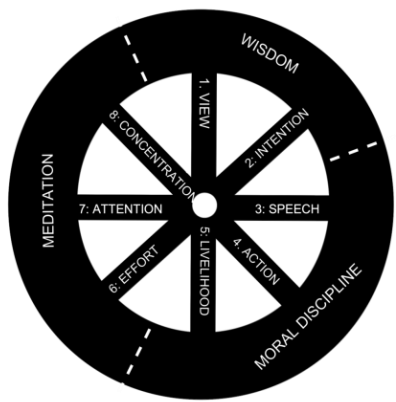

*Fig 1. Dhammacakka: the wheel of the teaching*

The eight factors of the path are not only intended to work together. They are also fundamentally integrated. For instance, in meditative training, the first aspect, immediately preceding mindfulness, is 'right effort'. It is traditionally described as giving rise to wholesome, or skillful (*kusala*) mental states not present yet, cultivating wholesome mental states already present, avoiding unwholesome, or unskillful (*akusala*) mental states not present yet, and diminishing unwholesome mental states already present. (Thanissaro, 1996; Adam, 2005) Right effort, therefore, is a monitoring kind of attention. Instead of being non-judgmental, it constantly evaluates experiences for their level of wholesomeness. Instead of accepting all, it is discerning, intent on actively building a character to embody an established ideal. And instead of being neutral, it distinguishes skilful from unskillful states. Right effort is blatantly moral to the point of the trainings of morality and meditation informing each other. Right effort also overlaps with the wisdom training, as wholesomeness is defined in terms of the Buddhist worldview ('right view') and its basic motivational attitude ('right resolve'), the latter consisting of raising a sustained interest in understanding the causes of suffering and a willingness to alleviate suffering in the world (Bombaerts et al., 2023). This is what we call 'embeddedness of attention' here: the fact that our attention is informed by our moral and philosophical frameworks that in turn are informed by our attention practices.

Our point is not so much that Buddhism is special because its attention practices are embedded, but that *any* kind of attention is embedded. Any kind of attention practice is linked to what is considered to be 'wholesome' action (the moral section of the wheel, fig.

1) and a set of adopted views on existence and basic intentions that lend direction to our lives for them to be sensed as meaningful (the wisdom section of the wheel, fig 1). A 'realist' and 'pragmatic' approach to the attention economy needs to take this embeddedness into consideration. We need to look at its participants not merely as users in an economic system, as Davenport and Beck do. Nor should we understand them merely as legal subjects with inalienable rights, as the liberal critique does. Rather, we must understand them as moral subjects in a shared world, within a 'social imagery', a 'socio-technical system' (McMahan, 2023; Hongladarom, 2020; Wehrle et al., 2022; Wellner 2022). And we must also understand that as individual users of the attention economy, our attentions, moral frameworks and worldviews are mutually constitutive and constantly in interaction with the attention economy and AI-systems in which we live, work, and consume.

### 3.2. Samyaksaṃkalpa: attending to the alleviation of suffering

*3.2.1 Desire-based attention versus commitment-based attention?*

Another Buddhist inspiration comes from comparing the function of desire in both systems. At face value, from the individual user's point of view, the contrast is simple: the attention economy's ideal user is desire-driven, whereas in Buddhism unskillful desire is a root cause of our suffering. Buddhism's aim is to deconstruct the false promises of what we think are our deepest desires and wake up to another less-desire based way of life. Buddhism indeed distinguishes non-systematic attention (*ayoniso-manasikara*), the so-called 'monkey-mind' on the one hand, and systematic, cultivated attention (*yoniso-manasikara*) on the other.The former just follows whatever desire crops up, the latter focuses on the aim of the Buddhist path as set out in the wisdom section of the eightfold path (Thanissaro 2006). Jay Garfield and Emily McRae (2024) call this second attention 'commitment-based' rather than desire-based.

Ironically – if not cynically – Davenport and Beck consider Buddhism as a possible ally of the attention economy in the sense that they express their hopes that Buddhist insight in the distracted, easily manipulated human attention can be put to use, so that "businesspeople who remember that every human brain is mainly a 'monkey mind' are ahead of others right off the mark." (2001, p. 71) This is in keeping with their self-image as realist and pragmatic thinkers rather than moralists.

Yet it is misleading to characterize the attention economy as desire-based and therefore non-committal. The name 'attention *economy'* may tempt us into thinking that users are merely engaged in an economic deal. But what we are actually buying (both in the sense of 'purchasing' and in 'accepting without further ado') is just as much an ideological ideal.

Depending on whether we approach the attention economy from the side of the producer (provider) or from the side of the consumer (user), two moral frameworks are at work here. For the provider, the moral framework is quite explicit: the resolve is to maximize profit, framed in what has been called 'the Californian Ideology': a mixture of leftist hippie freewheeling spirit and neoliberal yuppie entrepreneurial zeal and technological determinism (Barbrook and Cameron, 1996). On the users' side, the moral framework is far more covert, yet no less omnipresent and powerful. Tentatively and for brevity's sake we will describe it as *the free flow of easy gratification*. The implicit belief that all that is modern and good is merely a click away. It may sounds as a mere following of 'the monkey mind', but it presents itself as a morality in which effort, abstention and delayed desire are not worth pursuing. Just click and all will be well. This is not merely giving in to a non-concentrated wavering attention'. This is a systemic if implicit ideology and the way that this ideology is obscured could be considered an instance of epistemic injustice: it robs us not only of our attention, but also from seeing why we pay attention to the things that receive our attention.

When the liberal view criticizes the attention economy, it does so in the name of the same covert ideal: it blames the attention technology for not delivering its promise by meddling too much (e.g., the Onlife Manifesto (Floridi, 2015)). Users are alienated from their desires by technology creating demands where there were none. Or by subjecting them to data mining or abusing their private quests for fulfillment. Yet what is neglected in this view is the fact that in the attention economy the pursuit of desires through ease *is* its moral framework, an ideal that truly modern individuals are to be committed to, lest they want to run the risk of being considered outdated – a damning moral verdict.

Conversely it would be just as misleading to think of Buddhism as non-desire based. Pursuing the Buddhist ideal of awakening and liberation requires great and sustained motivation. In this respect some Buddhist authors (Sucitto, 2010) have distinguished two terms for desire. *Tanha* is used for the reflexive desire that chases things that seem to offer pleasantries, or do away with what promises to be unpleasant. *Chanda* on the other hand is used for the desire for sustainable happiness, related to a life committed to mindful, moral and wise practices. In this perspective, the reduction of Buddhist attention to non-desire based mindfulness rather leads to quietism and collaboration with prevalent systemic causes of suffering, rather than to taking a lively interest in the causes of suffering and in the active development of a better alternative.

*3.2.2. Self-committed attention vs attending to the suffering of the world*

Rather than opposing desire and commitment, the gap between the function of attention in the attention economy and in Buddhist practice is better described as a contrast between 'self-committed' versus 'liberation committed'. Both the attention economy and the liberal criticism present themselves as self-centered, in the sense that they claim to want each individual to live the life he or she intends to live. The liberal critique reproaches the attention economy of not delivering what it promises but does not question the promise itself.

Since its earliest days Buddhism presents itself explicitly as not self-based, to such an extent even that it is referred to as the teaching of non-self (*anatta*). Historically speaking this was a criticism of pre-Buddhist Indian mysticism, which claimed that liberation from suffering was to be found by meditating on one's true self or soul (*atta*) and thus awaken to one's oneness with the cosmic principle (Brahma), which was eternal, fully autonomous and ever blissful. By contrast the Buddha advised his followers to meditate on the observable experience that no such soul, eternity, autonomy or ever-lasting bliss was to be found (Gombrich, 2009, pp. 64-71). Rather than looking for liberation in a mystical Self, Buddhism looks for liberation in the cultivated attention of 'co-dependent arising' (*paticca-samuppada),* a radical sense of relationality in which to exist means to be related to myriads of causes and conditions, or even *to be* a cloud of interactions. That would be Buddhist 'realism', the first factor of the noble eightfold path, 'right view'. This relationality is taken as the basis for the second path factor, 'right resolve', which is not about finding one's true blissful Self, but about favoring wholesome (*kusala*) interactions over unwholesome (*akusala*) ones, in order to alleviate suffering in the world and enable a radically different way of being. That would be Buddhist 'pragmatics' (Hannes and Bombaerts, 2023).

Projected onto the criticism of the attention economy, the ideology of the free flow of easy gratification is closer to the mystical promise of a blissful Self than to Buddhist non-self-liberation. Buddhist resolve can serve as an inspiration for finding an alternative resolve, an alternative basic function of attention in our embedded lives. Here we propose to express this contrast as 'attention' versus 'attending'. As a noun, 'attention' can be seen as a resource to be exploited, or a currency one can pay with. Whereas 'attending' emphasizes cultivating a caring practice (Bombaerts et al., 2023). The Collins English Dictionary distinguishes eight current uses of the verb 'to attend', that can be grouped in three categories. One category simply refers to 'listening' or 'paying attention' (Collins, 3). A second one refers to what we have called an 'ecology': as being an integral part of a larger system (one attends classes or a church, Collins 1). Or as a phenomenon coming with or belonging causally to another phenomenon (Collins 4,5, a high fever attends a cold). A third category of meanings are associated with giving care (Collins 2), devoting oneself to something (Collins 6) or someone (Collins 2, 6-8), escorting (Collins 7) and providing for the needs (Collins 8) of somebody.

Historically speaking, Buddhism has emphasized the first person singular perspective, focusing on the cause of suffering on an individual level, even though some schools encourage to seek awakening for the benefit of all beings (Shi and Ewart, 2024). Philosophically speaking, 'liberation' can also apply to alleviating (individual or collective) suffering caused by systemic societal causes, as is the case for instance in the recent arising of engaged Buddhism. (Fuller, 2022) Beneficial interdependence can be cultivated on any organizational level. Sub-individual phenomena, like habits and patterns (Candiotto and Dreon, 2021), can be checked on their 'wholesomeness'. And so can super-individual systems like families, work floor cultures, political movements, classes, nations, international institutes and economical systems. We can substitute both the attention economy's view on society (as essentially an economy of self-interests) and the liberal view on society (as essentially a social contract between autonomous individuals) with a view on any societal level as a systemic ecology (Macy, 1991; Loy, 2019; Wirth, 2021) sharing attention practices, moral frameworks and views on existence.

In sum, if we are serious about offering an alternative to the AI-supported attention-economic system, we cannot limit ourselves to mildly defensive measures, in which we try to save some of the amount of attention available to us, take back what is robbed from us, or find ways to boost our attention again. We also need a fully worked out positive alternative approach to our attention. We need to be explicit about what alternative moral ideal we can be committed to in our attention.

In the end, it boils down to the seemingly simple question: *what are we doing when we are online*? Under the condition that we realize that our 'doing' is characterized by attentional, moral and philosophical practices, embedding our attention, practiced in a shared world, and that have us committed to ideals. Do we practice quietism in the face of systemic suffering, or do we develop forms of resistance based on a radically relational view aiming for beneficial relations in our societies?

It is our conviction that in order for us to answer these questions, we need to develop a solid view on both aspects we draw from basic Buddhist thinking: attention as an integral part of an 'ecology' and the function of attention to improve our attending to the suffering in the world.

## 4. A call for action

### 4.1. 'Ecology' as embedded attention: next steps

The Buddhist sources in section three can be of inspiration to change the context of the current attention economy's technology. It can call on us to consider individual users, designers, and big tech company employees and managers as moral subjects in a shared world whose attentions, moral frameworks and worldviews are mutually constitutive and constantly in interaction with the sociotechnical environment, such as AI systems. Several disciplines have already been developing notions of embedded attention (Doctor et al., 2022), but these notions remain scattered. Below we summarize several of these notions without a clear structure, which is, according to us, the current state of the field.

Crary's (1999) genealogical analysis traces the historical evolution of attention in Western capitalist societies. Since at least the nineteenth century, Western modernity has required individuals to define themselves by their ability to focus, naturalizing a form of human perceptual adaptability and continuously reshaping the boundaries between attention and distraction (Aloisi, 2023).

More recently, a "compass of modern attentional regimes" (Boullier, 2024; Citton, 2017; Campo, 2022) has been proposed to analyze different dynamics. The 'loyalty regime' calls for the users' commitment to stay tuned, to return to the social media platform, the smartphone or the app as often as possible. The 'alertness regime' installs an imperative to open to uncertainty and unpredictability. The 'projection regime' installs a sense of coherence into the observed world and entices us to ignore information that might contradict the framework. The 'immersion regime' brings users into a state of pleasurable fragmentation (Citton 2017, pp. 41-43). The political consequences of inhabiting different attentional regimes are significant, as projection regimes can become so prevalent that they shape our world without us being aware of them.

Changing the attention economy means intervening in this large historical evolution, which is *technologically* co-shaped. Puzio (2024) refers to "eco-relational" as the whole house we live in, which includes both "nature" and technologies. Humans are in a relationship with the whole non-human world. The notion of attention has been transformed over time, partly due to the technologies that shape it. In the age of the printed book, attention meant focusing on one activity at a time, and the background had to be silent and unnoticeable. In the age of radio and television, when stations and channels could be switched with a single button, attention was still focused on a single item but for shorter periods of time. Today, in the age of computers, cell phones, and AI, attention is distributed among tasks and appears in the form of "multi-attention (Wellner

2014; 2019; Gill, 2016, 2020). In its distributed form, attention can provide a subversive answer to the attention economy that requires our undivided attention.

A significant part of what we perceive is governed by social norms of relevance and irrelevance: precepts that determine what is worthy of attention ("social norms of focusing"), as well as norms that invite us to deliberately ignore things we notice (Zerubavel 2015, Campo 2022), or how we feel empathy (Concannon and Tomalin, 2023; Ho and Ho, 2024; Kleinrichert, 2024; Montemayor et al., 2022). We learn where to direct our attention through a process of socialization. As such, attention is embedded in our epistemological and moral context.

Concerning the epistemic embedding of attention, we refer to the work of Leonie Smith and Alfred Archer (2020) who coined the term *epistemic attention deficit* to refer to "to the unjust scenario in which someone is paid less *epistemic* attention than they *ought* to be paid." (2020, p. 779) They give the example of how newspapers or television shows can block certain perspectives, and hence from collective knowledge-production, for example by not interviewing refugees but only politicians debating migration. Epistemic attention deficits result from social power dynamics who accrue more importance to dominant perspectives than to marginalized ones. Becoming aware of this form of epistemic injustice requires that we realize which voices we want to hear, which perspectives we find worthy of our attention. Sometimes justice requires that we re-curb our attention, and unlearn habits (Schaubroeck, 2018). If social networks are epistemic environments (Blake-Turner, 2020), we need to ask ourselves what their heavy usage does to our epistemic attention and our related informational skills (Marin, 2022).

There is mutual influence of *morality and attention*. On the one hand, we direct our attention to the things that we find valuable, or at least worthy of our engagement. They matter, even when the value is negative (Watzl, 2022). On the other hand, how and whether we attend shapes our moral world-view, changes our moral quality, and enables or blocks our access to a reality outside ourselves. This second point is the insight brought to the ethics of attention, in different ways, by Iris Murdoch and Simone Weil. From the fact that "I can only choose in the world that I can see" (1970, p. 31), Murdoch derived that vision (or perception) rather than action was the moral faculty par excellence. Looking with just and loving attention is considered by Murdoch as 'the characteristic and proper mark of the active moral agent' (1998: 327). She stresses the ways in which ethics is bound to the individual's 'quality of consciousness', shaped by and shaping our attention. Like Murdoch, but more radically, Weil stresses the truth-seeking dimension of attention, as a way of giving our assent to a reality that we recognize as beyond our control, both epistemically and practically (1951, 1997). Both recommend training attention as an important part of (moral) education (Weil 1951, Murdoch 1992 ch.3), even regardless of the specific object of attention – which highlights the value they assign to attention as such (Caprioglio Panizza 2022).

In the interaction of individuals, attention in *economic, business, and governance performance* is not only considered as a commodity, but also as a commons that must be protected and cultivated alongside other efforts to dis-enclose the material and phenomenological commons (Doran, 2017; Brown and Zsolnai, 2018; Zsolnai, 2011). While it is well and good to bemoan the excesses of large technology companies, we need to seriously consider whether any realistic proposal to regulate the attention economy can actually produce the desired results. Governments and regulators face informational and motivational constraints paralleling those facing private actors. We should therefore be wary of assuming that even the best intended regulations will in fact achieve their aims (Chomanski, 2021). Using the framework of Non-Western Public Administration and Governance (Drechsler, 2015), current studies also have shed light on some of the governance practices in Asia that have historical Buddhist roots or are considered to be Buddhist and are relevant to the aimed research concerning attention economy and the Buddhist perspective towards it (Uttam, 2014; Shakya and Drechsler, 2019; Lennerfors, 2015).

In the *political* realm, Berger (2023) argues that the state of attention determines the efficacy of public spaces in articulating and achieving visions of the common good. As politicians seek to amass power by capturing attention, citizens can engage in disciplines of attention such as mindfulness in producing a public power that is more appropriately oriented to the welfare of all. These practices can enable enhanced ontological bonds to form between individuals, which can be the basis for more stable and effective political realities. Such bonds are not given but rather come about holistically by way of a hermeneutical circle of attention, language, and embodied understanding that recognizes the intrinsic value of mindfulness itself.

From a phenomenological perspective, attention is considered as an intensely *dynamic process* (Waldenfels, 2011; Arvidson, 2006; Losoncz, 2020). The specific feature of the role of attention is that it resides on a threshold of what is conscious/unconscious, known/unknown, within control/beyond control, visible/invisible, and is thus a dynamism that spans what are often considered mutually exclusive domains. In this sense, the dynamics of attention is what enables processes of change within the operational subject and various modes of attuning with the world and the other. Thus, attention becomes a key-faculty in the make-up of our psychological make-up, lived experience and, in extension, an existentially pivotal modality of our consciousness (Fredriksson 2022). Correlatively, the exploration and manipulation of its dynamism provide valuable insight into both the experienced dimension of our engagements with the world and others, as well as the possibilities to creatively reshape and enhance these engagements (Varela et al. 1991; Depraz et al. 2003).

Some researchers developed attention as grounded in the *body* through body sensations, physical movements, and interactions with our environment (D'Angelo, 2020; Marosan, 2023; Wehrle, 2020). We see the other's gaze and we respond to it with our gaze, looking at them in the eyes, as an embodied signal that we are minding them (Solomonova *et al.*, 2011). Attention is manifested as responsiveness - to the environment, to another's bid for attention, to artifacts and to ourselves - and this responsiveness is performed with our bodies: we raise our eyes from our smartphone's screen when something captures our attention, we look and gesture in particular ways when we are attentive to something or someone. The corporeal aspects of attention also bring current research in the fields of phenomenology and cognitive psychology into close contact with contemplative approaches within various "wisdom traditions." The latter have developed elaborate practical systems for exploring and modulating different dimensions of attention in concrete, hands-on settings (e.g., various meditative, breathing, somatic, etc., practices). It has therefore been suggested that a collaboration between scientific, philosophical, and contemplative approaches in the study of attention could prove methodologically, epistemologically and experimentally fruitful (Varela et al. 1991; Depraz *et al.*, 2023).

Meditation practitioners anchoring their attention on bodily sensations, breathing, and movement could detect mind wandering and regulate their attention and emotion (Kerr et al., 2013). Empirical research has been used to show that, by observing their bodily experiences without reaction, practitioners gain a deeper insight about the true characteristics of physical and mental phenomena (impermanent, suffering, not self) (Wongkitrungrueng and Juntongjin, 2022), and transform how they engage with pain and pleasure experiences. This embodied approach underscores the interconnectedness of mind and body in cultivating attention which contributes to the quality of attention as practitioners.

The above examples of notions of embedded attention are rich and promising, both at a theoretical and practical level. However, these efforts are still very scattered and lack synergy as a group. Future research should perform a continuous effort to bring these approaches in conversation in order to stimulate theoretical and policy-related synergies.

### 4.2. *Attending to the alleviation of suffering*

Once we acknowledge that there is a moral direction underlying the attention economy - so prevalent that it is near-invisible - we can and should wonder whether this is the moral framework we want to be directing our attention, values, decisions, but also our societal structures, such as the technological environment that forms a formidable presence in our current lives (Citton, 2017; Campo, 2023; Park, 2008).

As discussed in the third section, Buddhism emphasizes the mutual interrelation of our attention and our moral frameworks. *Sati*, one of its words for attention, can be understood both as 'noticing what passes by' and 'remembering: keeping in mind what the whole Buddhist path is about' (Analayo, 2003). This does not only mean that Buddhist attention is practiced within the moral framework of 'awakening' and 'liberation'. It also means that cultivated attention practices have the power to affect practitioners' moral frameworks. Attention practices make us aware of our covert moral frameworks. They can develop mental space for creating a critical distance with respect to ailments in the moral framework we appear to be living by. And they can offer room for other frameworks to grow as a new support for our attention.

It has been argued (McMahan 2023) that what traditional Buddhism calls 'enlightenment' or 'liberation' is a process of deconstructing an existing moral framework as suffering-engendering, and substituting it (or 'awakening' to) another framework that is experienced more efficient for tackling the suffering cause by a particular way of looking at life. Even within Buddhist history the content of the view awakened to *paticca samupada* has changed over the centuries and over the diverse Buddhist schools (McMahan, 2008). It is beyond the scope of this short paper to go into these views, as, again, our aim is not to give the outlines of attention in Buddhism. Here is the place to wonder what this could imply for the 'attention technology' we are embedded in, and how to resist the harmful effects of the current attention technologies.

As Colin Campbell (2015) has argued, moral frameworks are quite resistant to change, let alone to being substituted. They react like beings with a strong survival instinct. Even merely scrutinizing one's moral framework is hard and awkward work, accompanied with strong resistances. Yet we argue that we are in urgent need to do so with respect to the attention economy, and that attention researchers are well placed for doing so and for bringing the criticism of the attention economy to a higher and much needed level.

Three major questions can guide us in this.

1. How can technology help us to acknowledge that we are 'ecological' beings, rather than create the image of a self-sufficient, self-driven and self-serving individuals?

2. How can technology help us to raise awareness of our invariably being committed to a moral framework?

3. And if technologies are part of our being and our becoming and shape our moral decisions, the question arises: What kinds of technologies do we want? How should we design them? How can technology help us to raise the motivation for 'attending' better to the alleviation of suffering, by substituting our covert self-driven moral framework with an ecologically attending one? Not only to attend better to macro-problems such as climate change, concentration of power and wealth, and political polarization, but also, and perhaps even in the first instance, to attend better to that which the attention economy promises but fails to deliver: autonomy, the expression of our authenticity and finding contentment in life.

These questions are not only interesting to ethicists. We argue that they reflect a societal urgency. And the answer to them is that we do not really know.

Therefore, in the spirit of the manifesto-character of this paper, we call for all stakeholders in sociotechnical attention systems to collaborate. It would be exaggerated to state that "we have nothing to lose but our chains" (Marx and Engels, 1888), but we do have chains to pay attention to, to investigate them for their suffering-engendering characteristics and for hijacking not only our hormonal balance but also our moral frameworks and our societal embedding. A relevant technological environment is to be attuned to our long-term needs (Hershock, 2023; Hongladarom, 2020), as attending ecological beings, enabling us to cultivate freedom within our fully embedded existence (Williams, 2018).

## 5. Conclusion

The abstract of this paper served as a manifesto, a call for a shift from the current attention economy to an ecology of attending. By this we mean an approach that conceptualizes, designs and uses attention in an embedded and liberation-committed way. We turned to Buddhism for clarifying both aspects, yet we hope it is clear that, next to Buddhism and Buddhist ethics (Edelglass, 2013), we think that many other approaches can support our endeavor (e.g. Faruque, 2024; Depraz, 2003; Hiniker & Wobbrock, 2022; Weil, 1997).

This article hopes to be a stepping stone, offering a rough sketch of many potential conceptualizations that are to be elaborated and complemented with many others. On the basis of our reconceptualization of attention, attention economy applications can be redesigned, underlying business models further developed, and policy strategies worked out. New AI supported developments are inherently linked to attention, and here too the inevitable embeddedness of attention and a desirable liberation-commitment of attention should be built in the design of the digital applications.

We acknowledge that it is far from easy to come to grasps with what the above means concretely for the development of design. But we argue that this only increases the urgency of further research. Or the urgency of assembling all stakeholders to the table in order to discuss the future of embedded and liberation-committed attention in practice. To transform the inhumane socio-technical system of the attention economy into an ecology of attending.

**References**

Adam, M. T. (2005). Groundwork for a Metaphysic of Buddhist Morals: A New Analysis of puñña and kusala, in light of sukka. *Journal of Buddhist Ethics*, *12*, 62-85.

Aloisi, A. (2023). The Power of Distraction. https://www.torrossa.com/en/resources/an/5605849

Barbrook, R. and Cameron 1 (1996) The Californian Ideology. *Science as Culture* 6 (1), 44–72.

Berger, L. (2023). *The Politics of Attention and the Promise of Mindfulness*. Rowman & Littlefield.

Blake-Turner, C. (2020). Fake news, relevant alternatives, and the degradation of our epistemic environment. *Inquiry*, 1–21. https://doi.org/10.1080/0020174X.2020.1725623

Bombaerts, G., Anderson, J., Dennis, M., Gerola, A., Frank, L., Hannes, T., Hopster, J., Marin, L., & Spahn, A. (2023). Attention as Practice: Buddhist Ethics Responses to Persuasive Technologies. *Global Philosophy*, *33*(2), 25.

Boullier, D. (2024). 'The Drift of Attention Regimes in the Age of Digital Platforms'. *The Politics of Curiosity*, by E. Campo and Y. Citton, 116–27. Routledge. https://doi.org/10.4324/9781003401575-10.

Brown, C., & Zsolnai, L. (2018). Buddhist economics: An overview. *Society and Economy*, *40*(4), 497–513. https://doi.org/10.1556/204.2018.40.4.2

Campo, E. (2022). *Attention and its crisis in digital society*. Routledge. https://doi.org/10.4324/9781003225522

Campo, E. (2023). Exploring the Alternatives to the Attention Economy. *Contemporary Sociology*, *52*(5), 404–409. https://doi.org/10.1177/00943061231191420c

Candiotto, L., & Dreon, R. (2021). Affective scaffoldings as habits: A pragmatist approach. *Frontiers in psychology*, *12*, 629046.

Caprioglio Panizza, S. (2022). *Engaging the Real with Iris Murdoch and Simone Weil*. Routledge. https://doi.org/10.4324/9781003164852

Chomanski, B. (2021). The missing ingredient in the case for regulating big tech. *Minds and Machines*, *31*(2), 257-275.

Citton, Y. (2017). *The ecology of attention*. John Wiley & Sons.

Collins (n.d.). Attend. collinsdictionary.com. Retrieved September 10, 2024, from https://www.collinsdictionary.com/dictionary/english/attend.

Concannon, S., & Tomalin, M. (2023). Measuring perceived empathy in dialogue systems. *AI & SOCIETY*. https://doi.org/10.1007/s00146-023-01715-z

Crary, J. (1999). *Suspensions of Perception: Attention, Spectacle, and Modern Culture*. MIT Press.

D'Angelo, D. (2020). The phenomenology of embodied attention. *Phenomenology and the Cognitive Sciences*, *19*(5), 961–978. https://doi.org/10.1007/s11097-019-09637-2

Davenport, T., & Beck, J. (2001). *The attention economy: Understanding the new currency of business*. Harvard Business School Press.

Depraz, N. (2003). Between Christianity and Buddhism: Towards a Phenomenology of the Body–Mind. *Diogenes*, *50*(4), 23–32. https://doi.org/10.1177/03921921030504004

Depraz, N., Petitmengin, C., & Bitbol, M. (2023). Phenomenology and Mindfulness-Awareness. In *The Routledge Handbook of Phenomenology of Mindfulness* (pp. 491–501). Routledge. https://www.taylorfrancis.com/chapters/edit/10.4324/9781003350668-41/phenomenology-mindfulness-awareness-natalie-depraz-claire-petitmengin-michel-bitbol

Doctor, T., Witkowski, O., Solomonova, E., Duane, B., & Levin, M. (2022). Biology, Buddhism, and AI: Care as the driver of intelligence. *Entropy*, *24*(5), 710.

Doran, P. (2017). *A Political Economy of Attention, Mindfulness and Consumerism Reclaiming the Mindful Commons*. Routledge.

Drechsler, W. (2019). The Reality and Diversity of Buddhist Economics. *American Journal of Economics and Sociology*, 78(2), 523-560. https://doi.org/10.1111/ajes.12271

European Parliament. (2023). European Parliament resolution of 12 December 2023 on the addictive design of online services and consumer protection in the EU single market (2023/2043(INI)).

Edelglass, W. (2013). Buddhist ethics and western moral philosophy. *A companion to Buddhist philosophy*, 476-490.

Faruque, M. (2024). Attention, Consciousness, and Self-Cultivation in Sufi-Philosophical Thought. *Tasavvuf Araştırmaları Enstitüsü Dergisi*, *2*(2), Article 2. https://doi.org/10.32739/ustad.2023.4.56

Floridi, L. (2015). *The onlife manifesto: Being human in a hyperconnected era* (p. 264). Springer nature.

Fredriksson, A., & Panizza, S. (2022). Ethical attention and the self in Iris Murdoch and Maurice Merleau-Ponty. *Journal of the British Society for Phenomenology*, *53*(1), 24–39.

Fuller, P. (2022). An Introduction to Engaged Buddhism. Bloomsbury Academic.

Garfield, J.L. & McRae, E. (2024). The Wind in the Sails: Commitment in Śāntideva's How to Lead an Awakened Life. Harris, S. and P. Schmidt-Leukel (Eds). *Śāntideva and the Dynamics of Tradition: Doctrinal, Social, and Interreligious Contexts*. Bloomsbury.

Gill, K. S. (2016). Artificial super intelligence: Beyond rhetoric. *AI & SOCIETY*, *31*(2), 137–143. https://doi.org/10.1007/s00146-016-0651-x

Gill, K. S. (2020). Ethics of engagement. *AI & SOCIETY*, *35*(4), 783–793. https://doi.org/10.1007/s00146-020-01079-8

Gombrich, R. (2009), *What the Buddha Thought*. Sheffield: Equinox.

Hadar, L. L., & Ergas, O. (2019). Cultivating mindfulness through technology in higher education: A Buberian perspective. *AI & SOCIETY*, *34*(1), 99–107. https://doi.org/10.1007/s00146-018-0794-z

Hannes, T. & Bombaerts, G. (2023). What does it mean that all is aflame? Non-axial Buddhist inspiration for an Anthropocene ontology. *The Anthropocene Review*, https://doi.org/10.1177/2053019623115392.

Hannes, T. & Bombaerts, G. (2024) Being Taken for a Ride: Social and Technological Externalist Complements to the Internalist Reading of Buddhist Chariot Similes. *Philosophy East and West*, vol. 74 no. 3, 2024, p. 379-398. https://dx.doi.org/10.1353/pew.2024.a939596.

Hermann, E. (2023). Psychological targeting: Nudge or boost to foster mindful and sustainable consumption? *AI & SOCIETY*, *38*(2), 961–962. https://doi.org/10.1007/s00146-022-01403-4

Hershock, P. D. (2021). *Buddhism and intelligent technology: Toward a more humane future*. Bloomsbury Publishing.

Hershock, P. D. (2023). *Consciousness Mattering: A Buddhist Synthesis*. Bloomsbury Publishing. https://books.google.com/books?hl=en&lr=&id=oa3ZEAAAQBAJ&oi=fnd&pg=PP1&dq=hershock+consciousness+mattering&ots=4mUla9aH2j&sig=18LO0guC3gU_Fk1iUJ3Q1cNtzb8

Hiniker, A., & Wobbrock, J. O. (2022). *Reclaiming Attention: A Christian Perspective Prioritizing Relationships in the Design of Technology*.

Ho, M.-T., & Ho, M.-T. (2024). Bridging divides: Empathy-augmenting technologies and cultural soul-searching. *AI & SOCIETY*. https://doi.org/10.1007/s00146-024-01887-2

Hongladarom, S. (2011). Personal Identity and the Self in the Online and Offline World. *Minds and Machines*, *21*(4), 533–548. https://doi.org/10.1007/s11023-011-9255-x

Hongladarom, S. (2020). *The Ethics of AI and Robotics: A Buddhist Viewpoint*. Rowman & Littlefield.

Jacobs, B. (2017). *The original Buddhist psychology: What the Abhidharma tells us about how we think, feel, and experience life*. North Atlantic Books.

Kerr, C. E., Sacchet, M. D., Lazar, S. W., Moore, C. I., & Jones, S. R. (2013). Mindfulness starts with the body: Somatosensory attention and top-down modulation of cortical alpha rhythms in mindfulness meditation. *Frontiers in Human Neuroscience*, 7, 12. https://doi.org/10.3389/fnhum.2013.00012

Kleinrichert, D. (2024). Empathy: An ethical consideration of AI & others in the workplace. *AI & SOCIETY*. https://doi.org/10.1007/s00146-023-01831-w

Lennerfors, T. T. (2015). A Buddhist future for capitalism? Revising Buddhist economics for the era of light capitalism. *Futures*, *68*, 67–75.

Lin, C. T. (2023). All about the human: A Buddhist take on AI ethics. *Business Ethics, the Environment & Responsibility*, *32*(3), 1113-1122.

Losoncz, M. (2020). Concept of Attention in Bergson's and Husserl's Philosophy. *Filozofska Istrazivanja*, *40*(4), 829–839. https://doi.org/10.21464/fi40410

Loy, D. (2019). *Ecodharma: Buddhist teachings for the ecological crisis*. Simon and Schuster.

Macy, J (1991). *Mutual causality in Buddhism and general systems theory. The Dharma of natural systems.* State University of New York Press.

Marin, L. (2022). How to do things with information online. A conceptual framework for evaluating social networking platforms as epistemic environments. *Philosophy & Technology*, *35*(3), 77.

Marosan, B. P. (2023). The genesis of the minimal mind: elements of a phenomenological and functional account. *Phenomenology and the Cognitive Sciences*, 1-31.

Marx, K. & Engels, F. (1888) The Communist Manifesto [From the English edition of 1888, edited by Friedrich Engels], section IV. Retrieved Sept 14, 2024 from https://www.gutenberg.org/cache/epub/61/pg61-images.html.

Mascarello, C. (2018). Self-awareness in Tibetan buddhism. A study of the philosophical relevance of rang rig and its contribution to the contemporary debates on the nature of consciousness.

McGuire, B. F. (2016). Integrating the intangibles into asynchronous online instruction: Strategies for improving interaction and social presence. *Journal of Effective Teaching*, *16*(3), 62-75.

McMahan, D. L. (2008). *The making of Buddhist modernism*. Oxford University Press.

McMahan, D. L. (2023). *Rethinking Meditation: Buddhist Meditative Practice in Ancient and Modern Worlds.* Oxford University Press.

Montemayor, C., Halpern, J., & Fairweather, A. (2022). In principle obstacles for empathic AI: Why we can't replace human empathy in healthcare. *AI & SOCIETY*, *37*(4), 1353–1359. https://doi.org/10.1007/s00146-021-01230-z

Murdoch, I. (1992) *Metaphysics as a Guide to Morals*. London, Penguin.

Murdoch, I. (1998) *Existentialists and Mystics: Writings on Philosophy and Literature*. Edited by Peter Conradi. Harmondsworth, Penguin.

Murdoch, I. (1970/2001). *The sovereignty of good*. Routledge.

Park, J. Y. (2008). *Buddhism and postmodernity: Zen, Huayan, and the possibility of Buddhist postmodern ethics*. Lexington Books.

Puzio, A. (2024) Not Relational Enough? Towards an Eco-Relational Approach in Robot Ethics. *Philos. Technol.* **37**, 45. https://doi.org/10.1007/s13347-024-00730-2

Schaubroeck, K. (2018). Morality and what's love got to do with it. In *New Interdisciplinary Landscapes in Morality and Emotion*. Routledge.

Shakya, S., & Drechsler, W. (2019). The Guthis: Buddhist societal organization for the 21st century. Buddhism around the World. United Nations Day of Vesak 2019, 501-527.

Shi, J. & Ewart, G. (2024). A Humanistic Buddhist response to Crises through Mettā Verses. Shi, J.,Hill, S. Franzway, S. (Eds). *Cultivating compassion. Going beyond crises.* Peter Lang. Open access: https://www.peterlang.com/document/1340151.

Solomonova, E., Frantova, E., & Nielsen, T. (2011). Felt presence: The uncanny encounters with the numinous Other. *AI & SOCIETY*, *26*(2), 171–178. https://doi.org/10.1007/s00146-010-0299-x

Thanissaro (1996) *Magga-vibhanga Sutta: An Analysis of the Path* (SN 45.8). Retrieved Sept 3, 2024 from https://www.accesstoinsight.org/tipitaka/sn/sn45/sn45.008.than.html.

Thanissaro (2006) Untangling the Present, The Role of Appropriate Attention. Retrieved Oct 21, 2024 from https://www.accesstoinsight.org/lib/authors/thanissaro/untangling.html.

Tormen, F. (2023). Buddhism Has Always Been Posthuman: Philosophical Contributions to the Transhumanist Project. *Canadian Journal of Buddhist Studies*, (18).

Uttam, J. (2014). *The political economy of Korea: transition, transformation and turnaround*. Springer.

Van Vugt, M. K., Pollock, J., Johnson, B., Gyatso, K., Norbu, N., Lodroe, T., ... & Fresco, D. M. (2020). Inter-brain synchronization in the practice of Tibetan monastic debate. *Mindfulness*, *11*, 1105-1119.

Varela F.J., Evan, T., & Eleanor, R. (1991). *The embodied mind: Cognitive science and human experience*. Cambridge, Mass.: MIT Press.

Vörös, S. (2016). Sitting with the demons–mindfulness, suffering, and existential transformation. *Asian Studies*, *4*(2), 59-83.

Watzl, S. (2022) The Ethics of Attention: an argument and a framework. Archer, S. (ed.), *Salience: A Philosophical Inquiry*. New York, NY: Routledge.

Weele, C. V. D. (2022). Chapter 11 How Can Attention Seeking Be Good? From Strategic Ignorance to Self-Experiments. M. Wehrle, D. D'Angelo, & E. Solomonova (Eds). *Access and Mediation* (pp. 259–278). De Gruyter. https://doi.org/10.1515/9783110647242-012

Wehrle, M. (2020). Being a body and having a body. The twofold temporality of embodied intentionality. *Phenomenology and the Cognitive Sciences*, *19*(3), 499–521. https://doi.org/10.1007/s11097-019-09610-z

Wehrle, M., D'Angelo, D., & Solomonova, E. (Eds.). (2022). *Access and Mediation: Transdisciplinary Perspectives on Attention*. De Gruyter. https://doi.org/10.1515/9783110647242

Weil, S. (1951) Reflections on the Right Use of School Studies with a View to the Love of God. *Waiting for God,* translated by Emma Craufurd. New York, Harper and Row, pp. 105–116.

Weil, S. (1997). *Gravity and grace*. University of Nebraska Press.

Wellner, G. (2014). Multi-Attention and the Horcrux Logic: Justifications for Talking on the Cell Phone While Driving. *Techné: Research in Philosophy & Technology*, 18 (1-2) 48-73.

Wellner, G. (2019). Onlife Attention: Attention in the Digital Age. K. Otrel-Cass (Ed.) *Hyperconnectivity and Digital Reality: Towards the Eutopia of Being Human* (pp 47-65).

Wellner, G. (2022). Chapter 10 Attention and Technology: From Focusing to Multiple Attentions. M. Wehrle, D. D'Angelo, & E. Solomonova (Eds.), *Access and Mediation* (pp. 239–258). De Gruyter. https://doi.org/10.1515/9783110647242-011

Williams, J. (2018). *Stand out of our light: Freedom and resistance in the attention economy*. Cambridge University Press.

Wirth, J. (2021). Deep Social Ecology. *The Trumpeter*, *37*(1), 2-21.

Wongkitrungrueng, A., & Juntongjin, P. (2022). The Path of 'No' Resistance to Temptation: Lessons Learned from Active Buddhist Consumers in Thailand. *Religions*, *13*(8), Article 8. https://doi.org/10.3390/rel13080742

Zerubavel, E (2015). *Hidden in Plain Sight: The Social Structure of Irrelevance*. New York, NY: Oxford University Press.

Zheng, Y. (2024). Buddhist Transformation in the Digital Age: AI (Artificial Intelligence) and Humanistic Buddhism. *Religions*, *15*(1), 79.

Zsolnai, L. (2011). Why Buddhist Economics? L. Zsolnai (Ed.), *Ethical Principles and Economic Transformation—A Buddhist Approach* (pp. 3–17). Springer Netherlands. https://doi.org/10.1007/978-90-481-9310-3_1